\documentclass[11pt,preprint]{aastex}
\usepackage{lscape}
\begin{document}

\title{Looking outside the Galaxy:\\
the discovery of chemical anomalies in 3 old Large Magellanic Cloud 
clusters\footnote{Based on 
observations collected at the Very Large Telescope of the
 European Southern Observatory
(ESO), Cerro Paranal, Chile, under programme 080.D-0368(A).}}

\author{Alessio Mucciarelli}
\affil{Dipartimento di Astronomia, Universit\`a 
degli Studi di Bologna, Via Ranzani, 1 - 40127
Bologna, ITALY}
\email{alessio.mucciarelli2@unibo.it}

\author{Livia Origlia}
\affil{INAF - Osservatorio Astronomico di Bologna, Via Ranzani, 1 - 40127
Bologna, ITALY}
\email{livia.origlia@oabo.inaf.it}

\author{ Francesco R. Ferraro}
\affil{Dipartimento di Astronomia, Universit\`a 
degli Studi di Bologna, Via Ranzani, 1 - 40127
Bologna, ITALY}
\email{francesco.ferraro3@unibo.it}

\author{Elena Pancino}
\affil{INAF - Osservatorio Astronomico di Bologna, Via Ranzani, 1 - 40127
Bologna, ITALY}
\email{elena.pancino@oabo.inaf.it}

\begin{abstract}

By using the multifiber spectrograph FLAMES mounted at the ESO-VLT, 
we have obtained high-resolution spectra for 18 giant stars, 
belonging to 3 old globular clusters of the Large Magellanic 
Cloud (namely NGC~1786, 2210 and 2257). 
While stars in each cluster showed quite homogeneous iron 
content, within a few cents of dex (the mean values being 
Fe/H]=~--1.75$\pm$0.01 dex, --1.65$\pm$0.02 dex and --1.95$\pm$0.02 dex for 
NGC~1786, 2210 and 2257, respectively), we have detected significant 
inhomogeneities for the 
[Na/Fe], [Al/Fe], [O/Fe] and [Mg/Fe] abundance ratios, with 
evidence of [O/Fe] vs [Na/Fe]  and [Mg/Fe] vs [Al/Fe] 
anticorrelations.  The trends detected nicely agree with 
those observed in Galactic Globular Clusters, suggesting 
that such abundance anomalies are ubiquitous features of old 
stellar systems and they do not depend on the parent galaxy 
environment.
In NGC~1786 we also detected two extreme O-poor, Na-rich stars.
This is the first time that a firm signature of extreme chemical 
abundance anomalies 
has been found in an extragalactic stellar cluster.

\end{abstract}  
 

\section{Introduction}

Stars in Galactic Globular Clusters (GCs) are known to be homogenous 
in their overall metal content. However, in the recent years
significant star-to-star abundance variations of some light elements 
(namely Li, C, N, O, Na, Mg, Al) have been discovered 
\citep[see e.g.][]{gratton04, carretta08p, yong}, suggesting 
that Galactic GCs harbor at least 
two sub-populations with different chemical abundance ratios. 
The {\sl normal} population is similar to 
the halo field one, with average 
[O/Fe],[Mg/Fe] and [Al/Fe] enhanced by a factor 2-3 with respect to the 
solar value, and [Na/Fe] between solar and half-solar. The {\sl anomalous} 
population has significant [O/Fe] and moderate [Mg/Fe] depletion, and 
significant (up to an order of magnitude) [Na/Fe] and [Al/Fe] enhancement 
with respect to the solar value.
Moreover, super-O-poor stars  in the [O/Fe] distributions 
have been measured in a few Galactic GCs, as in 
NGC 2808 \citep{carretta06} and M13 \citep{sneden04}, reaching
[O/Fe]$\sim$~--1.2 dex, while [Na/Fe] remains about 
constant.

The detection of abundance anomalies in extragalactic clusters 
is still controversial and only few and sparce measurements exist so far.
\citet{hill00} analysed 3 giants of the old cluster NGC~2210 in the 
Large Magellanic Cloud (LMC), 
finding a general homogeneity for [O/Fe] (~+0.14 dex, $\sigma$=~0.10 dex) 
but a large spread for the [Al/Fe] ratio, 
with [Al/Fe]=~+0.11 dex ($\sigma$=~0.32 dex).
\citet{johnson06} discussed the chemical content of 4 LMC old clusters
(namely, Hodge 11, NGC~1898, 2005 and 2019). 
All the stars analysed have normal [O/Fe] and [Na/Fe]: only 
one star in NGC~1898 shows a largely enhanced [Al/Fe]=~+0.77 dex.
\citet{let} reported the chemical composition of 3 old clusters in Fornax, 
finding two stars (one in Fornax~1 and one in Fornax~3) with 
[O/Fe]$\sim$~+0.2 dex and enhanced [Na/Fe] $\sim$~+0.4 dex.
These measurements, although too sparce to probe any scenario, suggest that 
chemical anomalies can be found also in old extragalactic globulars.

This Letter is the third, after Ferraro et al.(2006) and 
Mucciarelli et al.(2008), of a series of papers aimed at studying 
in an homogeneous way the chemical composition of a sample of 
LMC clusters. For all the targets 
the same analysis procedure and basic assumptions for model 
atmospheres, atomic data and solar values have been adopted.
In this Letter we present chemical abundance patterns for 18 giants in 3 
old clusters in the LMC (namely NGC~1786, 2210 and 2257), with the purpose 
of clarifying the level of chemical inhomogeneity 
within these stellar systems.
The entire set of abundances for these clusters 
will be presented in a forthcoming paper.

\section{Observations and chemical analysis}

\begin{figure}
\plotone{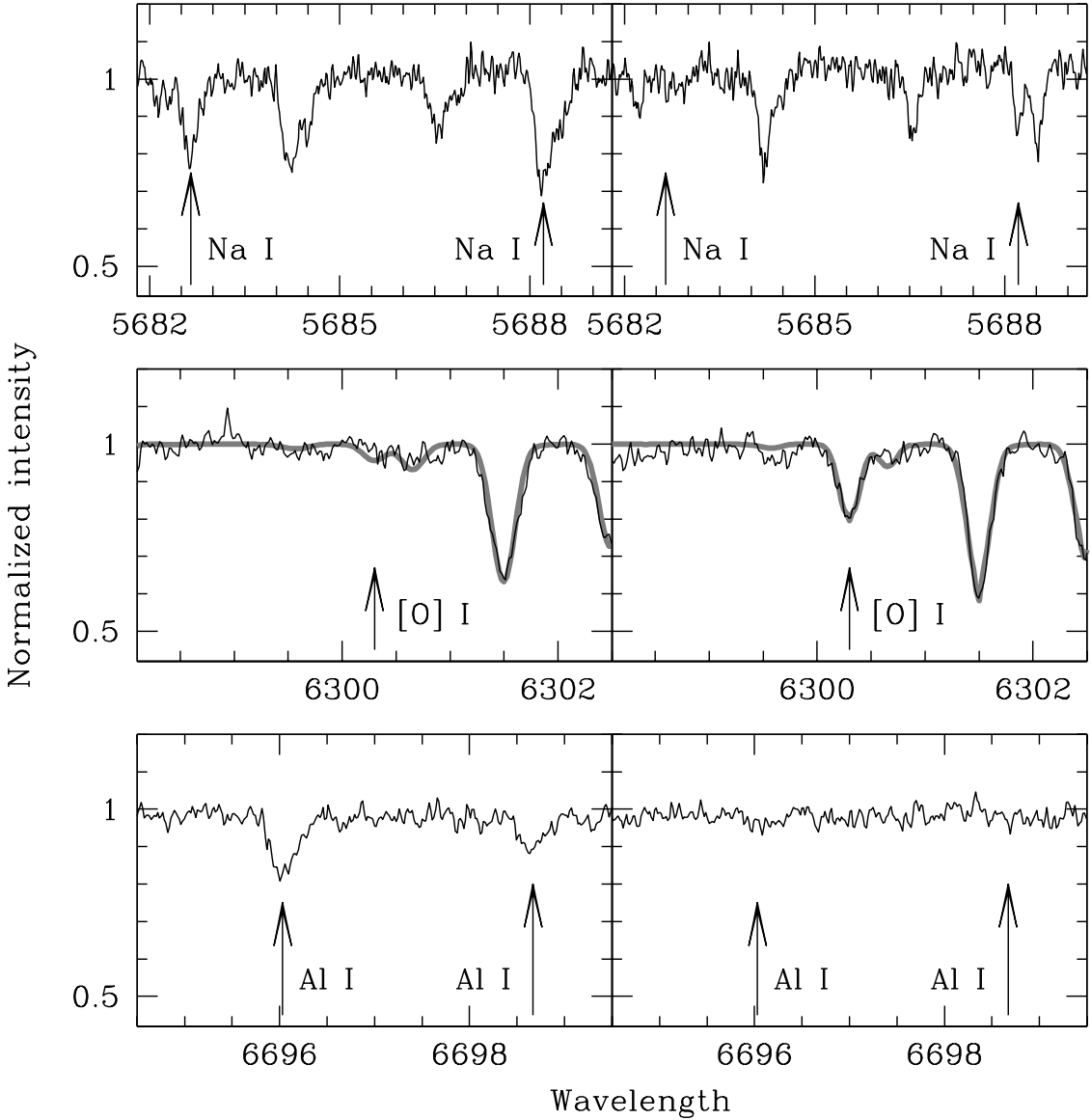}
\caption{Comparison between the spectra of the NGC~1786-2310 (left panels) 
and NGC~1786-1321 (right panels) stars. Arrows mark some Na, O and Al 
analysed features. The difference in the depth of these lines (despite 
similar atmospherical parameters) points toward real abundance differences 
in these stars. Moreover we plotted for the [O]~I lines also the 
corresponding best-fit synthetic spectra, computed with [O/Fe]=~-0.6 dex for 
the star NGC~1786-2310 (upper limit) and with [O/Fe]=~+0.25 dex for 
the star NGC~1786-1321.}
\label{spec}
\end{figure}

18 giants located in 3 old LMC clusters 
(NGC~1786, 2210 and 2257) have been observed
with FLAMES at the ESO-VLT telescope 
(ID Proposal 080.D-0368(A)).
The target stars have been selected along the Red Giant Branch 
of IR Color-Magnitude Diagram (A. Mucciarelli et al., 2009, 
in preparation) with typical 
$K_{0}\sim$12.8--13.5 magnitudes and 
$(J-K)_{0}\sim$0.6--0.8 colors.
All the targets have been observed with the UVES Red Arm 
(covering the $\sim$4800-6800 
$\mathring{A}$ wavelength range) and the HR~11 and HR~13 
GIRAFFE/MEDUSA setups (covering the 5600-5840 and 6120-6400 $\mathring{A}$ 
ranges, respectively).
The spectra have been acquired in series of 8-9 
exposures of $\sim$45 min each and pre-reduced by using the UVES and 
GIRAFFE ESO pipeline 
\footnote{http://www.eso.org/sci/data-processing/software/pipelines/}, 
including bias subtraction, flat-fielding, 
wavelength calibration and spectrum extraction. 
Single spectra have been coadded (reaching S/N$\sim$40-60 and 
S/N$\sim$80-100 per pixel for the UVES and GIRAFFE spectra, 
respectively) and finally normalized to unity in the continuum. 
Radial velocities have been measured by using the IRAF task 
FXCOR, performing a cross-correlation between the targets and 
a template of similar spectral type, and 
heliocentric corrections have been applied.
Fig.~\ref{spec} shows portions of the UVES spectra 
for two stars of NGC~1786, with some lines of interest 
indicated.

The abundance analysis has been performed using the LTE spectral line 
analysis code ROSA (developed 
by Raffaele G. Gratton, private communication) and  
the model atmosphere grids of \citet{kur}. 
The abundances of Fe, Na, Mg and Al have been computed from equivalent widths (EWs) 
measurements, while the O abundance has been obtained from spectral synthesis, 
to account for the contribution of the Ni line at 6300.34 $\mathring{A}$ 
to the line profile of the [O]~I line at 6300.31 $\mathring{A}$ 
(for sake of comparison, when possible, we 
measured also the  weaker [O]~I line at 6363.8 $\mathring{A}$).  
Na abundances have been derived from the 5682-88 
$\mathring{A}$ doublet and, when possible, the 6154-60 $\mathring{A}$ doublet, 
Mg has been measured from the 5711 and 6318 $\mathring{A}$ lines, 
while Al from the 6696-6698 $\mathring{A}$ doublet. 
O, Na and Mg lines are measured both in UVES and GIRAFFE 
spectra, while the Al lines are available only in the UVES spectra.
Oscillator strengths and atomic data for the selected 
lines are as in \citet{gratton03}, including those transitions 
for which accurate laboratory or theoretical log~gf are available, 
with formal errors below 0.05 dex.
Van der Waals broadening constants are from \citet{barklem00} and
the classical \citet{unsold} approximation.
Table~\ref{tab1} lists the wavelength, atomic parameters and measured 
EW for some lines, the complete version being available in electronic form.
Adopted solar values
are as in \citet{gratton03} who performed an analysis on the solar 
spectrum by using the same 
line-list and methodology adopted here.
Corrections for LTE departure have been applied to the Na abundances, 
using those computed by \citet{gratton99}.
For some stars the Al or [O]~I lines are too faint
and we provide upper limits only, derived 
by using synthetic spectra (for [O]~I) or 
(in the case of Al) computing the abundance corresponding 
to the minimum measurable EW. 
Guess values for $T_{eff}$ and log~g have been derived from dereddened 
JK photometry, by using the E(B-V) values reported by \citet{persson} 
and the $(J-K)_0$-$T_{eff}$ relation by 
\citet{alonso}.
Generally, the photometric $T_{eff}$ 
well satisfies the excitation equilibrium and only for a few targets we 
re-adjusted the $T_{eff}$ values in order to take into account 
residual trends between neutral iron abundances and excitation potential. 
Photometric gravities 
have been computed by using the isochrones of the Pisa Evolutionary Library 
\citep{cariulo} and adopting a distance modulus of $(m-M)_0$=~18.5 and an 
evolutionary mass of 0.85 $M_{\odot}$.
We re-adjusted also log~g values by imposing the ionization equilibrium 
between Fe I and Fe II.  
Also microturbulent velocities have been derived empirically by minimizing 
residual trends between Fe I abundances and expected line strengths.
For the GIRAFFE spectra, for which no Fe II line is available, 
gravities have been derived by using a suitable $T_{eff}$ vs log~g relationship, 
obtained from the UVES spectra.

\section{Results}

The average value of each abundance ratio (and the corresponding 
internal error and dispersion) for all the stars observed in each 
cluster is listed in Table~\ref{tab2}, together with 
the adopted atmospherical parameters.
Internal errors due to the EW measurement have been estimated 
by dividing the line-to-line scatter obtained in each star from the 
Fe I lines by the square root of the observed lines number. 
First guesses of the stellar parameters are from photometric 
information, then they are spectroscopically optimized. 
This means that the measurements of $T_{eff}$, log~g and $v_t$ 
are not independent. By varying each stellar parameter by a given 
amount and keeping the other fixed and
by summing in quadrature all these terms one can obtained an 
upper limit to the total uncertainty in the derived abundance.
To better estimate the uncertainty associated to the atmospherical 
parameters, we adopted the procedure already used by \citet{cayrel04} which 
take also into account the covariance terms. 
Since the largest uncertainties in the abundance determination arise 
from the uncertainty in $T_{eff}$,
we have varied the $T_{eff}$ of our best model by +100 K 
\footnote{This error corresponds to the 1$\sigma$ variation of
the slope between Fe~I abundances and excitation potential.}
and we repeated the analysis 
procedure for star NGC~1786-1501, chosen as 
representative of the entire sample, re-optimizing the other parameters.
This procedure provides variations of +0.08, +0.13, --0.07, --0.04 and 
--0.05 dex for [Fe/H], [O/Fe], [Na/Fe], [Mg/Fe] and [Al/Fe], respectively. 
The total error for each abundance ratio has been computed by the sum 
in quadrature of the errors due to EW measurement and atmospherical 
parameters.
In the following we schematically summarize the main results:\\ 

{\bf NGC 1786}: 
we measured seven stars, finding an iron content of [Fe/H]=~--1.75$\pm$0.01 
dex ($\sigma$=~0.02 dex) and an heliocentric radial velocity 
 $V_{r}$=~264.3 km $s^{-1}$ ($\sigma$=~5.7 km $s^{-1}$). 
 These results are consistent with the Ca II triplet analysis by \citet{ols} 
 that found [Fe/H]=~--1.87$\pm$0.2 dex.   
Two stars (\#2310 and \#2418) show large depletion 
of [O/Fe] and [Mg/Fe], and large enhancement of [Na/Fe] and 
[Al/Fe] (as measured in \#2310 only). 
Star \#1501 shows normal [O/Fe] and largely enhanced 
[Na/Fe] and [Al/Fe], while star \#978 show depletion 
of [O/Fe] and some enhancement of [Na/Fe].\\

{\bf NGC 2210}: 
we measured five giants, finding an iron content of 
[Fe/H]=~--1.65$\pm$0.02 dex ($\sigma$=~0.04 dex) and a mean heliocentric  
velocity of $V_{r}$=~337.5 km $s^{-1}$ ($\sigma$=~1.9 km $s^{-1}$). 
The previous analysis by \citet{ols} provides a lower iron content 
([Fe/H]=~--1.97$\pm$0.2 dex), while the high-resolution spectroscopic 
analysis by \citet{hill00} provides [Fe/H]=~--1.75$\pm$0.10 dex.
Both [O/Fe] and [Mg/Fe] appear normal and homogeneous, 
with average values of [O/Fe]=~+0.21 ($\sigma$=~0.09 dex) and 
[Mg/Fe]=~+0.33 ($\sigma$=~0.09 dex),
while [Na/Fe] and [Al/Fe] show a much larger spread. In particular,  
stars \#309 and \#431 have large ($>$0.5 dex) [Na/Fe] and [Al/Fe]
enhancement.\\

{\bf NGC 2257}: 
we measured six stars, finding an iron content of 
[Fe/H]=~--1.95$\pm$0.02 dex ($\sigma$=~0.04 dex) and heliocentric velocity 
of $V_{r}$=~299.4 km $s^{-1}$ ($\sigma$=~1.5 km $s^{-1}$). 
The Ca II triplet analysis performed by \citet{aaron} provides 
an higher iron content ([Fe/H]=~--1.59$\pm$0.02 dex), while 
the findings obtained by \citet{hill04} by using high-resolution 
spectroscopic point toward a lower value of iron ([Fe/H]=~--1.86$\pm$0.10 dex).
Four (namely \#189, \#586, \#842 and \#993) out of six stars are [O/Fe] depleted 
and [Na/Fe] enhanced. The other  two (\#136 and \#295) stars have normal [O/Fe] 
and enhanced [Na/Fe] and especially [Al/Fe] ($>$+0.8 dex) with 
respect to the solar value.\\

\begin{figure*}
\plottwo{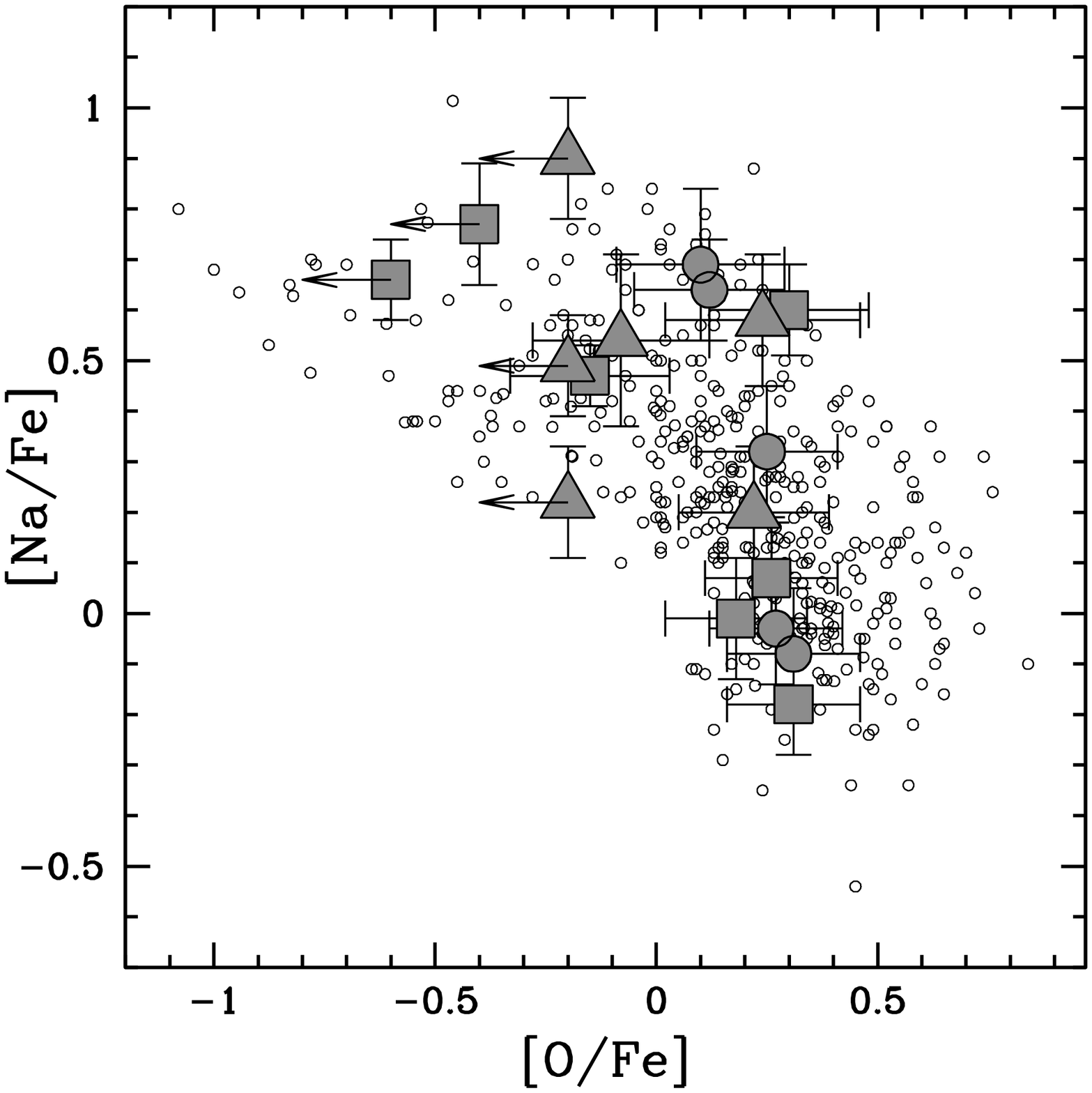}{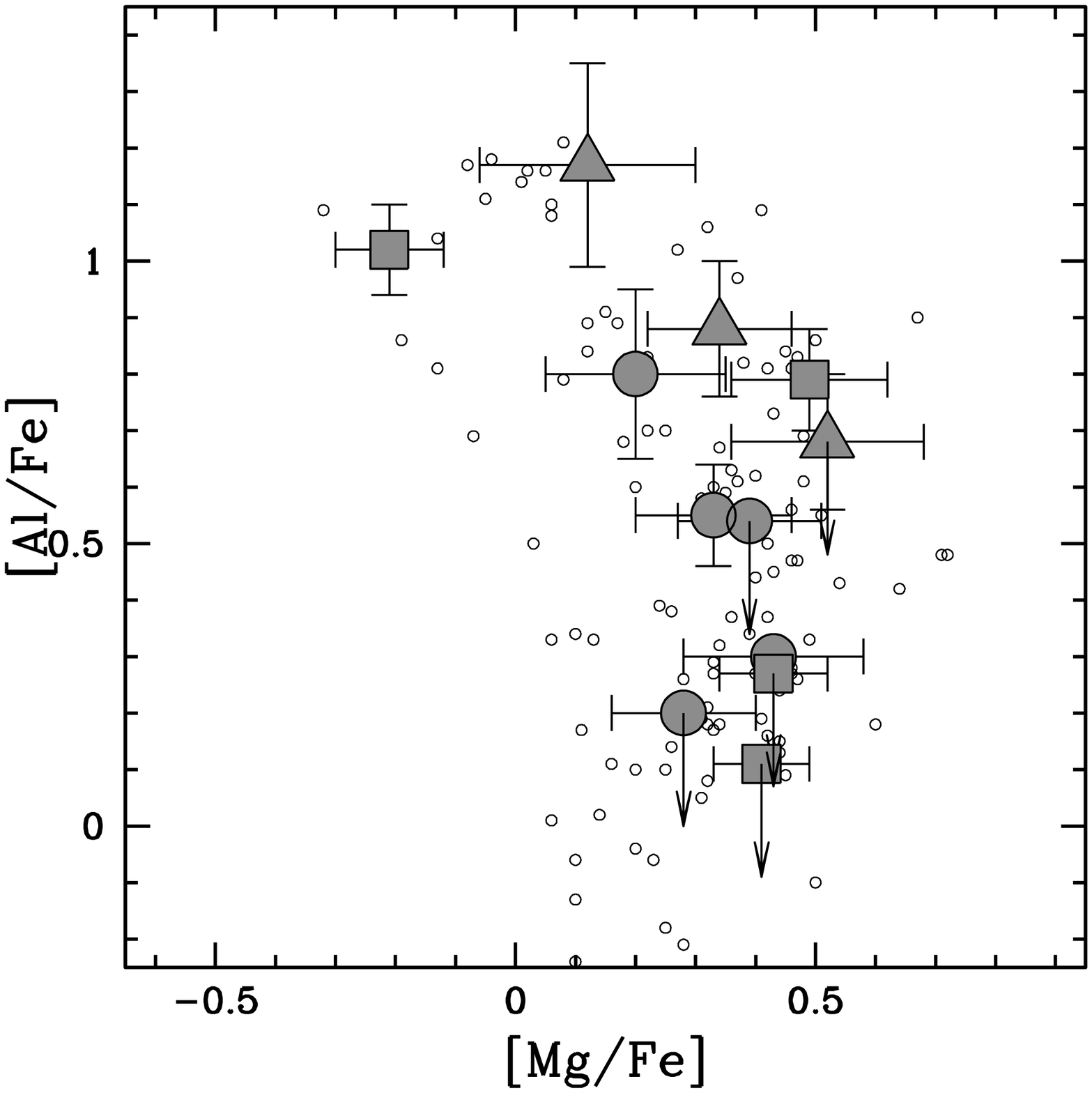}
\caption{Behaviour of [Na/Fe] as a function of [O/Fe] (left panel) 
and of [Al/Fe] as a function of [Mg/Fe] (right panel) for the target stars. 
Different symbols mark stars in different clusters: 
squares indicate the stars of NGC~1786, circles of NGC~2210, triangles of NGC~2257. 
Arrows indicate upper limits for the derived abundances.
For comparison small open circles are stars measured in Galactic Globular Clusters: 
M~3 \citep{sneden04}, M~4 \citep{ivans01}, M~5 \citep{ivans01}, M~13 \citep{sneden04}, 
M~15 \citep{sneden97}, M~71 \citep{ramirez02}
47 Tuc \citep{carretta04a}, NGC~2808 \citep{carretta04b, carretta06}, NGC~6218 \citep{carretta07b} 
and NGC~6752 \citep{gratton01, carretta07a}.
Error bars 
have been computed as the sum in quadrature of the uncertainty arising from 
EW measurements and atmospheric parameters.}
\label{ona}
\end{figure*}

The analysis presented in this Letter
demonstrates that the iron abundance  
(as well as the other iron peak and $\alpha$-elements, 
A. Mucciarelli et al., 2009, in preparation) is remarkably 
constant (within 0.1 dex) within each cluster. 
Conversely, the selected stars show a significant spread 
in the [Na/Fe] and [Al/Fe] abundance ratios, that 
largely exceed (at a level of 2-4~$\sigma$) the observational error. 
Moreover, in the case of NGC~1786 and NGC~2257 also a large spread 
in [O/Fe] and [Mg/Fe] has been found. 
On the other hand, the star-to-star differences in the absorption line 
intensities of Na, O and Al can be easily appreciated from the comparison 
shown in Fig.~\ref{spec}: this evidence confirms that a real spread in the 
content of these elements does exist among stars within the same cluster.
Fig.~\ref{ona} shows the behaviour of [Na/Fe] as a function of [O/Fe] 
(left panel) and of [Al/Fe] as a function of [Mg/Fe] (right panel) 
for the observed stars 
(stars in different clusters are marked with different symbols, 
see caption). 
As apparent from the Figure, the measured abundance ratios 
define a trend that nicely matches the distribution 
observed in Galactic GCs (small circles), tracing 
the so-called Na-O and Al-Mg anticorrelations. 
In order to quantify the correlation between the observed abundances,  
we measured the Spearman rank coefficient finding 
$C_S$=~--0.65 between [O/Fe] and [Na/Fe] 
and  $C_S$=~--0.50 between [Mg/Fe] and [Al/Fe], 
corresponding to a probability of $\sim$~99\% and $\sim$~95\%, 
respectively, that the two set of abundance ratios be anti-correlated.
{\it This is the first clear-cut evidence that Na-O and Al-Mg 
anticorrelations do occur in extragalactic clusters, suggesting 
that this feature might be ubiquitous to old globular clusters  
regardless of the parent galaxy environment}.

Fig.~\ref{ona2} shows the distribution of the [O/Fe] abundance ratio 
as a function of [O/Na] for the old LMC and Galactic GCs 
of Fig.~\ref{ona}. This plane is specially useful 
to trace the stars distribution along the Na-O anticorrelation and 
to better define the locus 
of the most extreme O-poor stars, defined as 
[O/Fe]$<$~--0.4 dex \citep{sneden04} and/or
[O/Na]$<$~--0.9 dex \citep{carretta08p2}. 
It is also worth noticing that two stars (\#2310 and \#2418) in 
NGC~1786 are super-O-poor (with [O/Fe]$<$--0.4 dex) 
and also largely depleted in [Mg/Fe] and (in the case of \#2310) 
largely enhanced in [Na/Fe] and [Al/Fe]. {\it This is the 
first time that an evidence of extreme chemical abundance anomalies 
has been firmly measured in an extragalactic stellar cluster}.

\section{Discussion and Conclusions}

The two main results of this Letter are: ({\it i})~the first discovery 
of Na-O and Mg-Al anticorrelations in the metal content of 
extragalactic GCs; ({\it ii})~the first detection 
of super-O-poor stars in such objects. 
Chemical anomalies and anticorrelations detected in Galactic 
GCs are currently interpreted in a scenario where the anomalous stars 
have formed from the ashes of a previous generation 
able to pollute the pristine gas with material processed by 
proton-capture reactions. Although still matter of debate, 
two possible polluters have been proposed:
intermediate-mass (3-8 $M_{\odot}$) AGB stars 
during the Hot Bottom Burning phase
\citep{ventura01, ventura08} and fast rotating 
massive stars \citep{decressin}. 
Theoretical models by \citet{dantona} suggest that 
the second generation of polluted stars should be a significant 
fraction of the entire cluster population ($\sim$50-70\%) and also 
He enriched. Recently, \citet{renzini08} proposed a critical 
discussion of the possible self-enrichment scenarios, showing as 
only the AGB stars are able to produce subsequent generations 
with the observed features.

Several cluster parameters affect the development and 
the extension 
of the anticorrelations, 
i.e. density, mass, metallicity, 
orbital parameters. Chemical anomalies have been detected 
exclusively in the Galactic GCs (with a clearcut correlation 
between cluster mass and the [O/Na] distribution) 
and not in the less massive and dense 
Open Clusters (OCs) or in the field stars, pointing out 
that the environment density and mass play a key role in retaining
the ejecta of the first generation.  
Less pronounced abundance anomalies at increasing cluster metallicity 
have been also suggested by theoretical models by \citet{ventura01}
and \citet{ventura08} in the framework of the AGB scenario. 
Moreover, the tight correlation showed by \citet{carretta06b} 
between the [O/Na] distribution and the orbital parameters of the 
Galactic GCs indicates that GCs less perturbed by the 
interactions with the Galactic Disk have more pronounced 
chemical anomalies.

\begin{figure}
\plotone{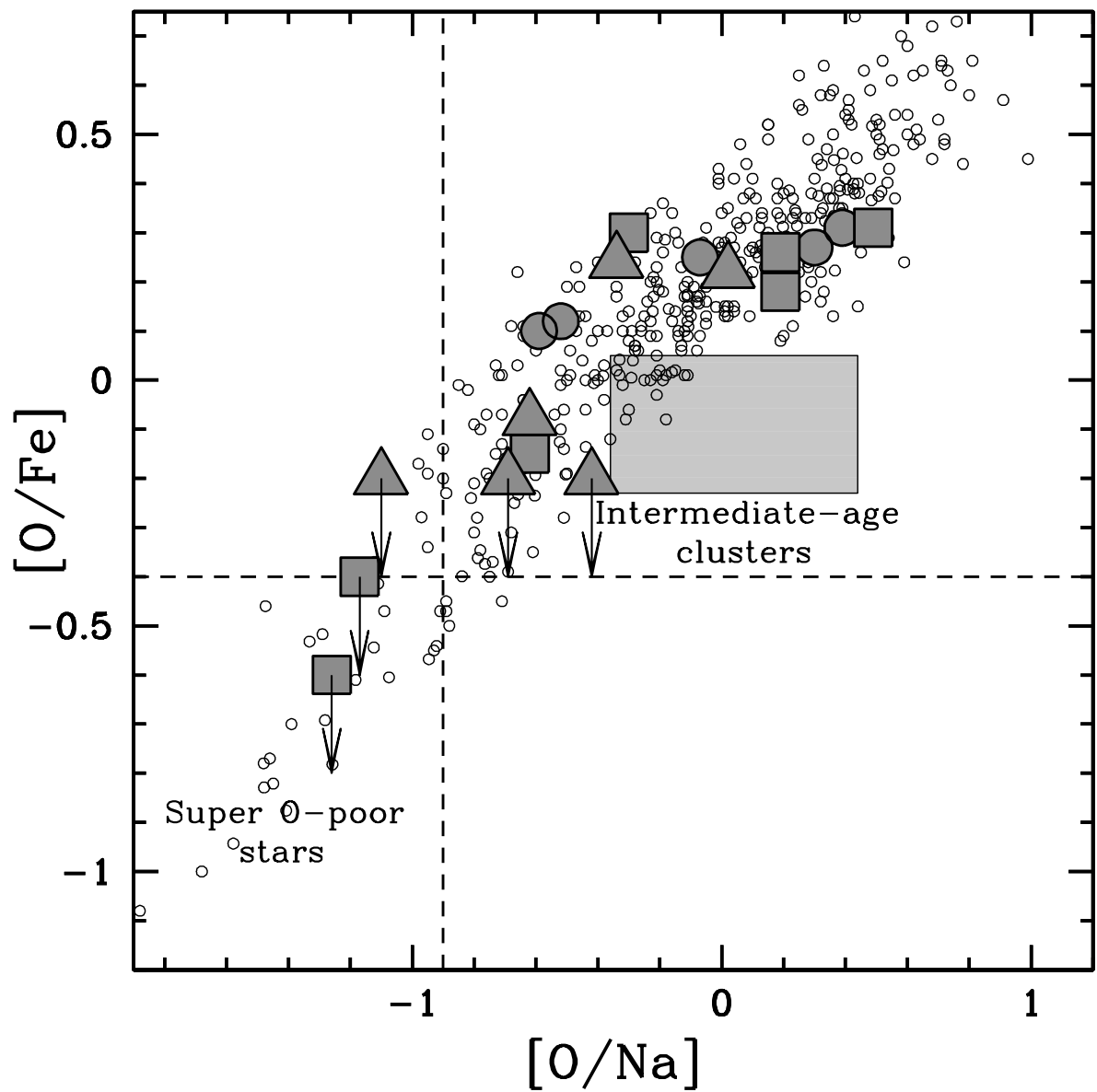}
\caption{ Behaviour of [O/Fe] as a function of [O/Na] for the target stars (same 
symbols as Fig.~\ref{ona}). Vertical and horizontal dashed lines mark the boundary 
of the super-O-poor stars defined by \citet{carretta08p2} and \citet{sneden04}. 
Grey box marks the position of the giant stars measured by \citet{m08}
in four intermediate-age LMC clusters.}
\label{ona2}
\end{figure}

The observations presented in this Letter demonstrate that this 
scenario can apply also to old clusters in the LMC. 
In Fig.~\ref{ona2}, for sake of comparison, we have also plotted 
(as a large grey rectangle)  the locus defined by 27 stars 
measured by \citet{m08} in 
four intermediate-age LMC clusters and obtained 
by analysing high-resolution spectra with the same 
methodology adopted here.
As apparent from the Figure, 
all the measured giants in the intermediate-age LMC clusters show 
[O/Na] values consistent 
with a {\sl normal} pre-enriched stellar population while 
[O/Fe] (for a given [O/Na]) is systematically lower than 
in old GCs (either Galactic or in the LMC). A similar 
(slightly subsolar) abundance has been found also for 
the other $\alpha$-elements abundances \citep{m08}. This is 
consistent with a pre-enrichment powered by both type II 
and Ia SNe over a long ($\sim$10 Gyr) timescale.

The old LMC clusters studied here show 
mass and metallicity 
comparable with the ones of the Galactic GCs.
The fact that, at variance with the older LMC and Galactic GCs, 
the intermediate-age LMC clusters do not show clear chemical 
abundance anomalies could be explained 
as a combined effect of high metallicity and moderate mass.
Indeed, the four intermediate-age clusters 
studied by \citet{m08} are significantly more metal-rich 
\footnote{However a quantitative comparison with the predictions of 
the models by \citet{ventura01}
and \citet{ventura08} cannot be performed since they provided  models 
up to a metallicity (Z=~0.004) which is too low to account for the 
intermediate-age LMC clusters (Z$\sim$0.008-0.009).}
and less massive (M$\sim5\cdot10^{4}$ to $\sim2\cdot10^{5} M_{\odot}$, 
Mackey \& Gilmore, 2003, D. Mackey, private communication) than 
the old LMC clusters studied here 
\citep[$\sim2-3\cdot10^{5} M_{\odot}$][]{mac}. 
It is worth noticing that these intermediate-age clusters are old enough 
to have already experienced self-enrichment (if any), because 
the pollution by AGB should work within the first 100 Myr of 
the cluster life.

The mass/metallicity threshold scenario also agrees with the fact that 
the few Galactic GCs with high metallicity ([Fe/H]$\sim$--0.7~/~--0.4 dex) 
and relevant abundance 
anomalies (as 47 Tuc, NGC~6388 and NGC~6441) 
are definitively the most massive clusters ($\sim 10^{6} M_{\odot}$)
in the Galactic system, while those with mass $\sim10^{5} M_{\odot}$
are all metal-poor.

The detection of abundance anomalies in the three old LMC 
clusters presented here suggests that old 
(relatively massive) metal-poor clusters can undergo  
self-enrichment process, regardless of the parent galaxy 
environment. The results presented here and in \citet{m08}, combined 
with the findings in the Galactic GCs and OCs, seem to suggest a 
mass/metallicity threshold effect that allows (if we consider 
the metal-rich regime) only very massive 
clusters to exhibit pronounced chemical anomalies. 
Moreover, due to not well known orbital parameters of the LMC 
clusters, we cannot exclude the effect due to the interaction with the 
Galaxy and SMC fields, similar to the one between several Galactic GCs 
and the Galactic Disk. 
\citet{bekki} point out that the LMC and SMC are in the present-day binary 
orbit in the last 5 Gyr, while 
\citet{besla} suggest as the LMC-SMC system is on its first 
close passage about the Galaxy, being entered in the Galactic virial 
radius less than $\sim$3 Gyr ago. 
These scenarios suggest as the most recent cluster populations 
could be strongly disturbed by the near tidal fields, both of the 
Galaxy and the SMC, pulling away the AGB ejecta and erasing
the chemical signatures of the self-enrichment 
also in the intermediate-age cluster 
massive enough. On the other hand,  
the old clusters have been less affected by similar perturbations, 
retaining the first-generation gas.
However, many more extragalactic clusters in the Magellanic Cloud 
and other Local Group galaxies need to be observed to further investigate 
the occurrence of this process with varying structural and physical 
cluster parameters and the host galaxy.

\acknowledgements  
We warmly thank the anonymous referee for his/her useful comments.
This research 
was supported by the  
Ministero dell'Istruzione, del\-l'Uni\-versit\`a e della Ricerca 
and it is part of the {\sl Progetti Strategici di Ateneo 2006} granted by the 
University of Bologna. The authors warmly thank Raffaele G. Gratton 
for useful suggestions in the abundance analysis procedure.

\begin{deluxetable}{cccccccccc} 
\tablecolumns{10} 
\tablewidth{0pc}  
\tablecaption{Atomic data and equivalent widths for the sample stars (in electronic form).}
\startdata 
 \hline
 \hline
   El. &   $\lambda$ & E.P. & log gf    & 1786-978      &       1786-1248   &     1786-1321  &  1786-1436&    1786-1501  & 1786-2310  \\
&   ($\mathring{A}$) &     (eV)    &  &  (m$\mathring{A}$) &   (m$\mathring{A}$) &  (m$\mathring{A}$) 
& (m$\mathring{A}$) & (m$\mathring{A}$) & (m$\mathring{A}$) \\
 \hline
 \hline  
    O I &  6300.31    & 0.00 &  -9.75    &   syn       &       syn      &	syn   &    syn  &     syn      &     syn    	           \\
   O I &  6363.79    & 0.02 & -10.25    &   syn       &       syn      &	syn   &    syn  &     syn      &     syn    	          \\
   Na I & 5682.65    & 2.10 &  -0.67    &   26.0      &  17.9	       &  10.0        &    0.0  &  47.8    &  58.2     \\
   Na I & 5688.22    & 2.10 &  -0.37    &   63.0      &  34.4	       &  24.2        &   22.2  &  77.0    &  78.0     \\
   Na I & 6154.23    & 2.10 &  -1.57    &    0.0      &  0.0	       &  0.0         &    0.0  &   9.3    &  0.0      \\
   Na I & 6160.75    & 2.10 &  -1.26    &   14.1      &  0.0           &  0.0	      &    0.0  &  16.6    &  25.5	 \\
   Mg I & 5711.09    & 4.34 &  -1.73    &   52.8      &  66.6          &  67.6        &   53.0  &  77.7    &  36.2  \\
   Mg I & 6318.71    & 5.11 &  -1.94    &    0.0      &  12.0          &  13.3        &    0.0  &   0.0	   &   0.0	       \\
\enddata   
\label{tab1}
\end{deluxetable}

\setlength{\topmargin}{3pt}
\setlength{\textheight}{24cm}

\begin{landscape}
\begin{deluxetable}{lcccccccccc}
\tablecolumns{11} 
\tiny
\tablewidth{0pt}
\tablecaption{Adopted atmospherical parameters and derived abundances for the NGC~1786, 2210 and 2257 LMC clusters. 
Last column indicates the corresponding adopted spectrum, G for GIRAFFE and U for UVES.}
\tablehead{ 
\colhead{ID} &    \colhead{$T_{eff}$} &
 \colhead{log~g}& 
   \colhead{[A/H]}  & 
   \colhead{$v_t$} 
 &  \colhead{[Fe/H]}& 
   \colhead{[O/Fe]} 
 &  \colhead{[Na/Fe]} &
   \colhead{[Mg/Fe]} 
 &  \colhead{[Al/Fe]} & \colhead{spectrum}
  \\
  &  \colhead{(K)} & & \colhead{(dex)} & \colhead{(km $s^{-1}$)} 
  & \colhead{(dex)} & \colhead{(dex)} &   \colhead{(dex)} & \colhead{(dex)} & \colhead{(dex)} &}
\startdata 
\hline
 SUN &        &       &         &     & 7.54  & 8.79 &    6.21      &   7.43 & 6.23  &  \\
\hline
 &        &       &         &     &   & NGC~1786 &          &    &   &  \\
\hline
978    &  4250  & 0.57  & --1.75  &  1.40 &  --1.73~$\pm$~0.02  &--0.15~$\pm$~0.12 & 0.47~$\pm$~0.03   &  0.25~$\pm$~0.06  & ---	   & G \\
1248   &  4280  & 0.75  & --1.75  &  1.70 &  --1.74~$\pm$~0.02  & 0.26~$\pm$~0.08  & 0.07~$\pm$~0.08   &  0.43~$\pm$~0.08  & $<$0.27	   & U\\
1321   &  4250  & 0.65  & --1.75  &  1.80 &  --1.73~$\pm$~0.01  & 0.31~$\pm$~0.07  &--0.18~$\pm$~0.07  &  0.41~$\pm$~0.07  & $<$0.11	   & U\\
1436   &  4420  & 0.76  & --1.75  &  1.70 &  --1.76~$\pm$~0.02  & 0.18~$\pm$~0.09  &--0.01~$\pm$~0.09  &  0.40~$\pm$~0.09  &  ---	   & G \\
1501   &  4100  & 0.55  & --1.80  &  1.80 &  --1.79~$\pm$~0.01  & 0.30~$\pm$~0.08  & 0.60~$\pm$~0.06   &  0.49~$\pm$~0.12  & 0.79~$\pm$~0.08	& U\\
2310   &  4100  & 0.47  & --1.75  &  1.90 &  --1.72~$\pm$~0.01  & $<$--0.60	   & 0.66~$\pm$~0.05   & --0.21~$\pm$~0.08 & 1.02~$\pm$~0.06	& U\\
2418   &  4160  & 0.47  & --1.80  &  1.50 &  --1.75~$\pm$~0.02  &$<$--0.40	   & 0.77~$\pm$~0.03   & --0.31~$\pm$~0.07 & ---		  & G \\  
\hline
Average         &        &       &         &       &  --1.75~$\pm$~0.01  &--0.01~$\pm$~0.14 & 0.35~$\pm$~0.14   &  0.22~$\pm$~0.13  & 0.55~$\pm$~0.16  &  \\
$\sigma$        &        &       &         &       &    0.02             & 0.37             & 0.36              &  0.34             & 0.43             &  \\
\hline 
   &        &       &         &  &  & NGC~2210             &    &   &  \\
\hline
122    &  4300  & 0.60  & --1.65  &  1.70 &  --1.66~$\pm$~0.02  & 0.31~$\pm$~0.08  &--0.08~$\pm$~0.11  &  0.39~$\pm$~0.11  & $<$0.54	   & U \\
309    &  4250  & 0.55  & --1.70  &  1.80 &  --1.69~$\pm$~0.03  & 0.10~$\pm$~0.14  & 0.69~$\pm$~0.10   &  0.20~$\pm$~0.14  & 0.80~$\pm$~0.14	& U \\
431    &  4200  & 0.70  & --1.65  &  1.80 &  --1.67~$\pm$~0.02  & 0.12~$\pm$~0.11  & 0.64~$\pm$~0.07   &  0.33~$\pm$~0.12  & 0.55~$\pm$~0.08	& U \\
764    &  4270  & 0.60  & --1.60  &  1.90 &  --1.58~$\pm$~0.02  & 0.25~$\pm$~0.10  & 0.32~$\pm$~0.10   &  0.43~$\pm$~0.14  & $<$0.30	   & U \\
1181   &  4200  & 0.60  & --1.60  &  1.80 &  --1.64~$\pm$~0.02  & 0.27~$\pm$~0.08  &--0.03~$\pm$~0.08  &  0.28~$\pm$~0.11  & $<$0.20	   & U \\
\hline
Average         &        &       &         &       &  --1.65~$\pm$~0.02  & 0.21~$\pm$~0.04  & 0.31~$\pm$~0.16   &  0.33~$\pm$~0.04  & 0.48~$\pm$~0.10  &  \\
$\sigma$        &        &       &         &       &    0.04             & 0.09             & 0.36              &  0.09             & 0.23             &  \\
\hline 
  &        &       &         &   &  &  NGC~2257         &  &    &   &  \\
\hline
 136	&  4290  & 0.65  & --1.90  &  1.95 &  --1.94~$\pm$~0.02  & 0.22~$\pm$~0.11  & 0.20~$\pm$~0.11	&  0.34~$\pm$~0.11  & 0.88~$\pm$~0.11	 & U \\
 189	&  4290  & 0.61  & --1.90  &  1.60 &  --1.92~$\pm$~0.02  & $<$--0.20	    & 0.49~$\pm$~0.07	&  0.42~$\pm$~0.10  &  ---	    & G \\
 295	&  4360  & 0.96  & --2.00  &  1.50 &  --1.95~$\pm$~0.03  & 0.24~$\pm$~0.18  & 0.58~$\pm$~0.10	&  0.12~$\pm$~0.18  & 1.17~$\pm$~0.18	 & U \\
 586	&  4480  & 0.82  & --2.00  &  1.50 &  --1.92~$\pm$~0.03  & $<$--0.20	    & 0.22~$\pm$~0.08	&  0.36~$\pm$~0.11  &  ---	    & G \\
 842	&  4320  & 0.95  & --1.90  &  1.50 &  --1.96~$\pm$~0.02  &--0.08~$\pm$~0.15 & 0.54~$\pm$~0.10	&  0.52~$\pm$~0.15  & $<$0.68	    & U \\
 993	&  4200  & 0.52  & --2.00  &  1.50 &  --2.02~$\pm$~0.03  & $<$--0.20	    & 0.90~$\pm$~0.09	&  0.24~$\pm$~0.13  &  ---	    & G \\
\hline
Average         &        &       &         &       &  --1.95~$\pm$~0.02  &--0.04~$\pm$~0.08 & 0.49~$\pm$~0.11   &  0.33~$\pm$~0.06  & 0.91~$\pm$~0.10  &  \\
$\sigma$        &        &       &         &       &    0.04             & 0.21             & 0.26              &  0.14             & 0.25             &  
\enddata   
\label{tab2}
\end{deluxetable}
\end{landscape}

\end{document}